\documentclass[aps,prd,onecolumn,showpacs,groupedaddress,nofootinbib]{revtex4}  
\usepackage{graphicx}
\usepackage{epstopdf}
\usepackage{amsmath}
\usepackage{amsfonts}
\usepackage{amssymb}
\usepackage{appendix}
\usepackage{comment}
\usepackage{bbold}
\usepackage{color}
\usepackage{slashed}
\usepackage{subfigure}
\usepackage{setspace}
\usepackage{footnote}
\usepackage{multirow}
\usepackage{braket}
\usepackage[normalem]{ulem}
\usepackage[a4paper, portrait, margin=0.5in]{geometry}
\usepackage[utf8]{inputenc}
\usepackage{mathrsfs}

\usepackage{url}
\usepackage{wrapfig}
\usepackage{braket}

\begin{document}
\singlespacing
{\hfill NUHEP-TH/16-03}

\title{False Signals of CP-Invariance Violation at DUNE}

\author{Andr\'{e} de Gouv\^{e}a}
\affiliation{Northwestern University, Department of Physics \& Astronomy, 2145 Sheridan Road, Evanston, IL 60208, USA}
\author{Kevin J. Kelly}
\affiliation{Northwestern University, Department of Physics \& Astronomy, 2145 Sheridan Road, Evanston, IL 60208, USA}

\begin{abstract}
One of the main goals of the Deep Underground Neutrino Experiment (DUNE) is to look for new sources of CP-invariance violation. Another is to significantly test the three-massive-neutrinos paradigm. Here, we show that there are CP-invariant new physics scenarios which, as far as DUNE data are concerned, cannot be distinguished from the three-massive-neutrinos paradigm with very large CP-invariance violating effects. We discuss examples with non-standard neutrino interactions and with a fourth neutrino mass eigenstate. We briefly discuss how ambiguities can be resolved by combining DUNE data with data from other long-baseline experiments, including Hyper-Kamiokande. 
\end{abstract}

\pacs{13.15.+g, 14.60.Pq, 14.60.St}
\maketitle

\section{Introduction}
\label{sec:Introduction}
Neutrino oscillations have been measured with remarkable precision and nearly all data collected agree with the three-massive-neutrinos paradigm, where the Standard Model Lagrangian is augmented to include nonzero neutrino masses, but no new accessible states or interactions are present. The next generation of oscillation experiments plans to precisely study the phenomenon of neutrino oscillations and test, nontrivially, the current paradigm. Among the highest priorities is the search for leptonic CP-invariance violation. The most comprehensive and ambitious next-generation projects are the Deep Underground Neutrino Experiment (DUNE)~\citep{Adams:2013qkq,Acciarri:2016crz,Acciarri:2015uup} and the Hyper-Kamokande experiment (HyperK) \cite{Kearns:2013lea,Abe:2014oxa,Abe:2015zbg}. 

In the three-neutrino paradigm, there is only one source of leptonic CP-invariance violation that is accessible to oscillation experiments: the Dirac CP-odd phase $\delta$ in the Pontecorvo-Maki-Nakagawa-Sakata (PMNS) leptonic mixing matrix. The current long-baseline oscillation experiments, Tokai to Kamiokande (T2K) \cite{Abe:2011ks} and the NuMI [Neutrinos at the Main Injector] Off-Axis $\nu_e$ Appearance Experiment (NO$\nu$A) \cite{Ayres:2004js} have begun to detect hints for a non-zero value of $\delta$ when their data \cite{Abe:2012gx,Abe:2013hdq,Adamson:2016tbq} are combined with results from the reactor experiments Double CHOOZ \cite{Abe:2012tg}, Daya Bay \cite{An:2012eh}, and RENO~\citep{Adamson:2016xxw}. More or less up-to-date combinations of the current neutrino data can be found in \cite{Forero:2014bxa,Gonzalez-Garcia:2015qrr,Capozzi:2016rtj,Agashe:2014kda}. If there is additional new physics in the lepton sector, however, new sources of CP-invariance violation are also expected to be present. Whether next-generation experiments are sensitive to these new sources is the subject of intense phenomenological research, and depends on the nature of the new phenomenon.

Here, we concentrate on a rather contrasting question. We investigate whether the presence of new CP-conserving physics may result in experiments measuring the value of $\delta\neq 0$, $\pi$ incorrectly, and hence concluding that leptons violate CP-invariance. Specifically, we concentrate on data consistent with what might be observed by the DUNE experiment, and consider the cases of non-standard neutrino neutral-current interactions (NSI) and a four-neutrino hypothesis. Both of these topics have been widely explored in the literature. NSI were first proposed as a solution to the solar neutrino problem~\cite{Wolfenstein:1977ue}, and they have been explored in depth for their effects on oscillations involving solar neutrinos~\citep{Krastev:1992zx,Miranda:2004nb,Bolanos:2008km,Palazzo:2009rb,Escrihuela:2009up,Friedland:2004pp}, atmospheric neutrinos~\citep{GonzalezGarcia:1998hj,Fornengo:1999zp,Fornengo:2001pm,Huber:2001zw,Friedland:2004ah,Friedland:2005vy,Yasuda:2010hw,GonzalezGarcia:2011my,Esmaili:2013fva,Choubey:2014iia,Mocioiu:2014gua,Fukasawa:2015jaa,Choubey:2015xha}, and accelerator neutrinos~\citep{Friedland:2006pi,Blennow:2007pu,EstebanPretel:2008qi,Kopp:2010qt,Coloma:2011rq,Friedland:2012tq,Coelho:2012bp,Adamson:2013ovz,Girardi:2014kca,deGouvea:2015ndi,Coloma:2015kiu,Blennow:2015nxa,Liao:2016hsa,Forero:2016cmb,Huitu:2016bmb,Bakhti:2016prn,Miranda:2016wdr,Coloma:2016gei}. A fourth neutrino has been proposed as a solution to the short-baseline anomalies~\citep{Aguilar:2001ty,AguilarArevalo:2008rc,Mention:2011rk,Frekers:2011zz,Aguilar-Arevalo:2013pmq} and its impact on oscillations has been studied both for short-baseline~\citep{deGouvea:2014aoa,Choubey:2016fpi,Giunti:2015wnd,Gariazzo:2015rra,Adey:2015iha} and long-baseline~\citep{Donini:2007yf,Dighe:2007uf,deGouvea:2008qk,Meloni:2010zr,Bhattacharya:2011ee,Hollander:2014iha,Klop:2014ima,Giunti:2015wnd,Gariazzo:2015rra,deGouvea:2015euy,Tabrizi:2015bba,Gandhi:2015xza,Palazzo:2015gja,Agarwalla:2016mrc,Agarwalla:2016xxa} experiments.

In general, both the NSI and four-neutrino scenarios accommodate additional sources of CP-invariance violation. Some NSI parameters are complex and mediate CP-invariance violating neutrino--matter interactions, while the addition of a fourth neutrino requires extending the PMNS matrix to an additional eigenstate; this opens up new sources of CP-invariance violation in the mixing matrix. Recent work has focused on parameter degeneracies in neutrino oscillations at long baselines (see, e.g., Refs.~\citep{Liao:2016hsa,Forero:2016cmb,Masud:2016bvp,Miranda:2016wdr,Coloma:2016gei}), particularly those involving NSI. Our current contribution differs from the existing literature by focusing on CP-conserving new neutrino physics that may be misinterpreted as CP-violating. 

This manuscript is organized as follows. In Sec.~\ref{sec:NSIand3P1}, we discuss, very briefly,  the formalism of neutrino oscillations in the three-neutrino paradigm, in the presence of  NSI, and in a four-neutrino scenario. In Sec.~\ref{sec:DUNE}, we introduce the DUNE experiment and the simulation we perform. In Sec.~\ref{sec:Results}, we explore the circumstances for fake CP-invariance violation, and in Sec.~\ref{sec:Conclusions}, we discuss, semi-qualitatively, how data from other oscillation experiments may aide in determining the true hypothesis, and offer concluding remarks.

\section{Oscillations, Non-Standard Interactions, and Four-Neutrino Scenarios}
\label{sec:NSIand3P1}

We direct the reader to Refs.~\citep{deGouvea:2015ndi,Coloma:2015kiu} and~\cite{Berryman:2015nua} for more comprehensive discussions of non-standard neutrino interactions (NSI) and four-neutrino scenarios in the context of long baseline neutrino experiments, respectively. Here, we briefly introduce the three-neutrino oscillation formalism, and in Secs.~\ref{subsec:NSI} and~\ref{subsec:4Nu}, we introduce the notation used for NSI and four-neutrino scenarios.

Within the three-neutrino paradigm, flavor oscillations are parameterized by two linearly independent mass-squared differences $\Delta m_{ij}^2 \equiv m_j^2 - m_i^2$; $i,j=1,2,3$, where $m_i$ are the neutrino mass eigenvalues, and the four parameters that define the PMNS matrix: the three mixing angles $\theta_{12}$, $\theta_{13}$, and $\theta_{23}$, and the CP-odd phase $\delta$. Throughout, we use the Particle Data Group definition of the three-neutrino mixing parameters \cite{Agashe:2014kda}. Except for $\delta$, all these parameters have been measured with good precision \cite{Agashe:2014kda}; we discuss their values in more detail later. The probability for a neutrino flavor eigenstate $\ket{\nu_\alpha}$ to propagate a distance $L$ and be detected as a flavor eigenstate $\ket{\nu_\beta}$ ($\alpha,~\beta$ = $e,~\mu,~\tau$) is denoted as $P_{\alpha\beta} \equiv |\mathcal{A}_{\alpha\beta}|^2$, where 
\begin{equation}
\mathcal{A}_{\alpha\beta} = \bra{\nu_\beta} U e^{-i H_{ij} L} U^\dag \ket{\nu_\alpha},
\end{equation}
$U$ is the PMNS matrix (a function of the angles $\theta_{ij}$ and $\delta$), and $H_{ij}$ is the propagation Hamiltonian in the basis of the neutrino mass eigenstates. In the ultra relativistic approximation, and in the absence of interactions between the propagating neutrinos and the matter along the path of propagation, the Hamiltonian is diagonal and can be written as $H_{ij} = 1/ (2E_\nu)\; \mathrm{diag}\left\lbrace 0, \Delta m_{12}^2, \Delta m_{13}^2\right\rbrace$. Interactions with electrons, protons, and neutrons introduce an effective background potential, diagonal in the flavor eigenbasis, and
\begin{equation}
H_{ij} \to H_{ij} + U_{i\alpha}^\dag V_{\alpha\beta} U_{\beta j},
\end{equation}
where $V_{\alpha\beta}$ is the background potential. We can absorb the (diagonal) interactions with protons and neutrons, as well as the neutral-current interactions with the background electrons, as an overall phase in the amplitude, and all that remains is $V_{\alpha\beta} = A\; \mathrm{diag}\left\lbrace 1, 0, 0\right\rbrace$, where $A = \sqrt{2} G_F n_e$, and $G_F$ and $n_e$ are the Fermi constant and number density of electrons along the path of propagation, respectively. In the Earth's crust $A \simeq 10^{-4}$ eV$^2 /$GeV, a small parameter when compared to $|\Delta m^2_{13}|/E_{\nu}$ for $E_{\nu}\sim 1$~GeV. When considering oscillations involving antineutrinos, $A \to -A$ and $U \to U^*$. Even for $\delta = 0$ or $\pi$, when the weak interactions are CP-invariant in the lepton-sector, matter effects generically introduce an asymmetry between CP-conjugated oscillation probabilities, e.g., $P_{\mu e} \neq P_{\bar{\mu} \bar{e}}$ in matter.

\subsection{Non-Standard Neutrino Interactions}
\label{subsec:NSI}

We assume that the following effective Lagrangian mediates non-standard neutrino interactions with ordinary matter:
\begin{equation}
\mathcal{L}^{\mathrm{NSI}} = -2\sqrt{2} G_{F} \left(\bar{\nu}_\alpha \gamma_\rho \nu_\beta\right) \left( \epsilon_{\alpha\beta}^{f \tilde{f} L} \bar{f}_L \gamma^\rho \tilde{f}_L + \epsilon_{\alpha\beta}^{f \tilde{f} R} \bar{f}_R \gamma^\rho \tilde{f}_R \right) + h.c.,
\end{equation}
where $\epsilon_{\alpha\beta}^{f\tilde{f}s}$ characterize the strength, relative to the weak interactions, of the interaction between neutrinos of flavor $\alpha$ and $\beta$ with fermions $f_s$ and $\tilde{f}_s$ of chirality $s$. As in Refs.~\citep{deGouvea:2015ndi,Friedland:2004pp,Friedland:2004ah,Friedland:2005vy,Yasuda:2010hw,GonzalezGarcia:2011my,Choubey:2014iia,Friedland:2006pi,Ohlsson:2012kf,Kikuchi:2008vq}, we make the following assumptions. First, $f = \tilde{f}$ -- we only consider diagonal, neutral-current couplings to the charged fermions, and $f = e,~u,~d$, as we are concerned with neutrinos propagating through the Earth. Second, we only consider effects on neutrino propagation, and ignore contributions of NSI to the production or detection of neutrinos. We define $\epsilon_{\alpha \beta} \equiv \sum_f \epsilon_{\alpha\beta}^{f} n_f / n_e$, where $\epsilon_{\alpha\beta}^f \equiv \epsilon_{\alpha\beta}^{ffL} + \epsilon_{\alpha\beta}^{ffR}$, and $n_f$ is the number density of fermion $f$. For propagation through the Earth, we assume $n_u = n_d = 3n_e$. Under these assumptions, NSI effects amount to a modification of the matter background potential: 
\begin{equation}
V_{\alpha\beta} \longrightarrow A\left( \begin{array}{c c c} 1 + \epsilon_{ee} & \epsilon_{e\mu} & \epsilon_{e\tau} \\ \epsilon_{e\mu}^* & \epsilon_{\mu\mu} & \epsilon_{\mu\tau} \\ \epsilon_{e\tau}^* & \epsilon_{\mu\tau}^* & \epsilon_{\tau\tau} \end{array} \right).
\end{equation}
The NSI parameters $\epsilon_{\alpha\beta}$ are complex for $\alpha \neq \beta$, and each of these three parameters is described by a magnitude and a CP-odd phase $\phi_{\alpha\beta}$: $\epsilon_{\alpha\beta} \equiv |\epsilon_{\alpha\beta}| e^{i\phi_{\alpha\beta}}$.

All current experimental data are consistent with $\epsilon_{\alpha\beta} = 0$, and the following ``neutrino-only'' bounds at 90\% confidence level (CL) 
\begin{equation}
\label{NSIBounds}
\left( \begin{array}{c c c} |\epsilon_{ee}| < 4.2 & |\epsilon_{e\mu}| < 0.33 & |\epsilon_{e\tau}| < 3.0 \\ & |\epsilon_{\mu\mu}| < 0.07 & |\epsilon_{\mu\tau}| < 0.33 \\ & & |\epsilon_{\tau\tau}| < 21\end{array}\right)
\end{equation}
can be found in Ref.~\cite{Ohlsson:2012kf}. The bounds above are mostly independent of the complex phases $\phi_{\alpha\beta}$.

As explored, for example, in Ref.~\cite{deGouvea:2015ndi,Coloma:2015kiu}, the Deep Underground Neutrino Experiment is fertile ground for exploring NSI, and future DUNE data, if consistent with the three-neutrino paradigm, would in general improve bounds on the NSI parameters by an order of magnitude compared to those in Eq.~(\ref{NSIBounds}). One final remark: neutrino oscillation experiments are not sensitive to all three diagonal NSI parameters, as any term proportional to the identity can be absorbed as an overall phase when calculating $\mathcal{A}_{\alpha\beta}$. With this in mind, and taking advantage of the relatively strong bounds on $\epsilon_{\mu\mu}$, we operationally set $\epsilon_{\mu\mu}$ to zero. In practice, the distinction between an upper bound for $|\epsilon_{\alpha\alpha} - \epsilon_{\mu\mu}|$ and $|\epsilon_{\alpha\alpha}|$ for $\alpha = e,~\tau$ is not significant.

\subsection{Four-Neutrino Scenario}
\label{subsec:4Nu}

A fourth neutrino mass eigenstate would, in general, also have a nonzero probability of being detected as an active neutrino $\nu_{\alpha}$. One can accommodate this possibility with the introduction of new mass-squared differences $\Delta m_{i4}^2 \equiv m_4^2 - m_i^2$ ($i=1,2,3$) and the extension of the $U(3)$ PMNS matrix into a $U(4)$ matrix. Here, we restrict our discussion to $m_4 > m_1$ so $\Delta m_{14}^2 > 0$, however we make no such claim relating $m_2$ and $m_3$ to $m_4$. We require $U$ to be a unitary matrix, which can be parameterized, as far as oscillations are concerned, by six mixing angles and three phases. To reduce confusion between the three- and four-neutrino scenarios, we refer to these mixing angles as $\phi_{ij}$; $i < j$; $i,$ $j = 1$, $2$, $3$, $4$, and to the phases as $\eta_1$, $\eta_2$, and $\eta_3$. In the limit that $\phi_{i4} \to 0$, the three-neutrino mixing angles $\theta_{12,13,23}$ are equivalent to $\phi_{12,13,23}$ and the phase $\eta_1$ can be identified with the three-neutrino phase $\delta$. Concretely, we choose the relevant matrix elements to be
\begin{align}
U_{e2} =&~ s_{12} c_{13} c_{14}, \\
U_{e3} =&~ s_{13} c_{14} e^{-i \eta_1}, \label{eq:Ue3}\\
U_{e4} =&~ s_{14} e^{-i \eta_2}, \\
U_{\mu 2} =&~ c_{24} \left( c_{12} c_{23} - e^{i\eta_1} s_{12} s_{13} s_{23}\right) - e^{i(\eta_2 - \eta_3)} s_{12} c_{13} s_{14} s_{24}, \\
U_{\mu 3} =&~ s_{23} c_{13} c_{24} - e^{i(\eta_2 - \eta_3 - \eta_1)} s_{13} s_{14} s_{24}, \\
U_{\mu 4} =&~ s_{24} c_{14} e^{-i\eta_3}, \\
U_{\tau 2} =&~ c_{34} \left(-c_{12} s_{23} - e^{i\eta_1} s_{12} s_{13} c_{23}\right) - e^{i\eta_2} c_{13} s_{12} c_{24} s_{14} s_{34} -e^{i\eta_3} \left(c_{12} c_{23} - e^{i\eta_1} s_{12} s_{13} s_{23}\right) s_{24} s_{34}, \\
U_{\tau 3} =&~ c_{13} c_{23} c_{34} - e^{i(\eta_2 - \eta_1)} s_{13} c_{24} s_{14} s_{34}  - e^{i\eta_3} s_{23} c_{13} s_{24} s_{34}, \\
U_{\tau 4} =&~ c_{14} c_{24} s_{34},
\end{align}
where $s_{ij} \equiv \sin{\phi_{ij}}$ and $c_{ij} \equiv \cos{\phi_{ij}}$. This is identical to the parameterization in Ref.~\cite{Berryman:2015nua}. In practice, we will only be interested in $P_{\mu\mu}$ and $P_{\mu e}$ (plus the CP-conjugated channels). These oscillation probabilities are only sensitive to two of the three CP-odd phases. 

In four-neutrino scenarios, the propagation Hamiltonian must be modified in two ways. First, the neutrino Hilbert space is four-dimensional, and the ``kinetic-energy'' term becomes
\begin{equation}
H_{ij} \longrightarrow \frac{1}{2E_\nu} \mathrm{diag}\left\lbrace 0, \Delta m_{12}^2, \Delta m_{13}^2, \Delta m_{14}^2\right\rbrace,
\end{equation}
to account for the new mass eigenstate. Second, the enlarged matter potential can be written as
\begin{equation}
V_{\alpha\beta} \longrightarrow A \left( \begin{array}{c c c c} 1 & 0 & 0 & 0 \\ 0 & 0 & 0 & 0 \\ 0 & 0 & 0 & 0 \\ 0 & 0 & 0 & \frac{n_n}{2 n_e}\end{array}\right),
\end{equation}
where $n_n$ is the number density of neutrons. Throughout, as is customary, we assume $n_n = n_e$. This last point is related to the assumption -- which we make throughout -- that the fourth neutrino weak-eigenstate is sterile, i.e. it does not couple to the $W$-boson or the $Z$-boson. 

\section{The Deep Underground Neutrino Experiment}
\label{sec:DUNE}

The Deep Underground Neutrino Experiment (DUNE) is a proposed next-generation neutrino oscillation experiment with a conventional muon neutrino beam (or antineutrino beam) generated at Fermilab that propagates $1300$~km to a liquid argon time-projection chamber in South Dakota. We consider that DUNE consists of a $34$ kiloton detector and $1.2$ MW proton beam, consistent with the proposal in Ref.~\cite{Adams:2013qkq}. The neutrinos range in energy between $0.5$ and $20$ GeV with a peak around $3.0$ GeV. We consider six years of data collection: three years each with the neutrino and antineutrino beams. We consider that the sign of $\Delta m_{13}^2$, i.e.~the mass hierarchy, will be determined prior to DUNE collecting data, and restrict our analysis to the normal hierarchy, $\Delta m_{13}^2 > 0$.

\begin{figure}
\centering
\includegraphics[width=0.8\linewidth]{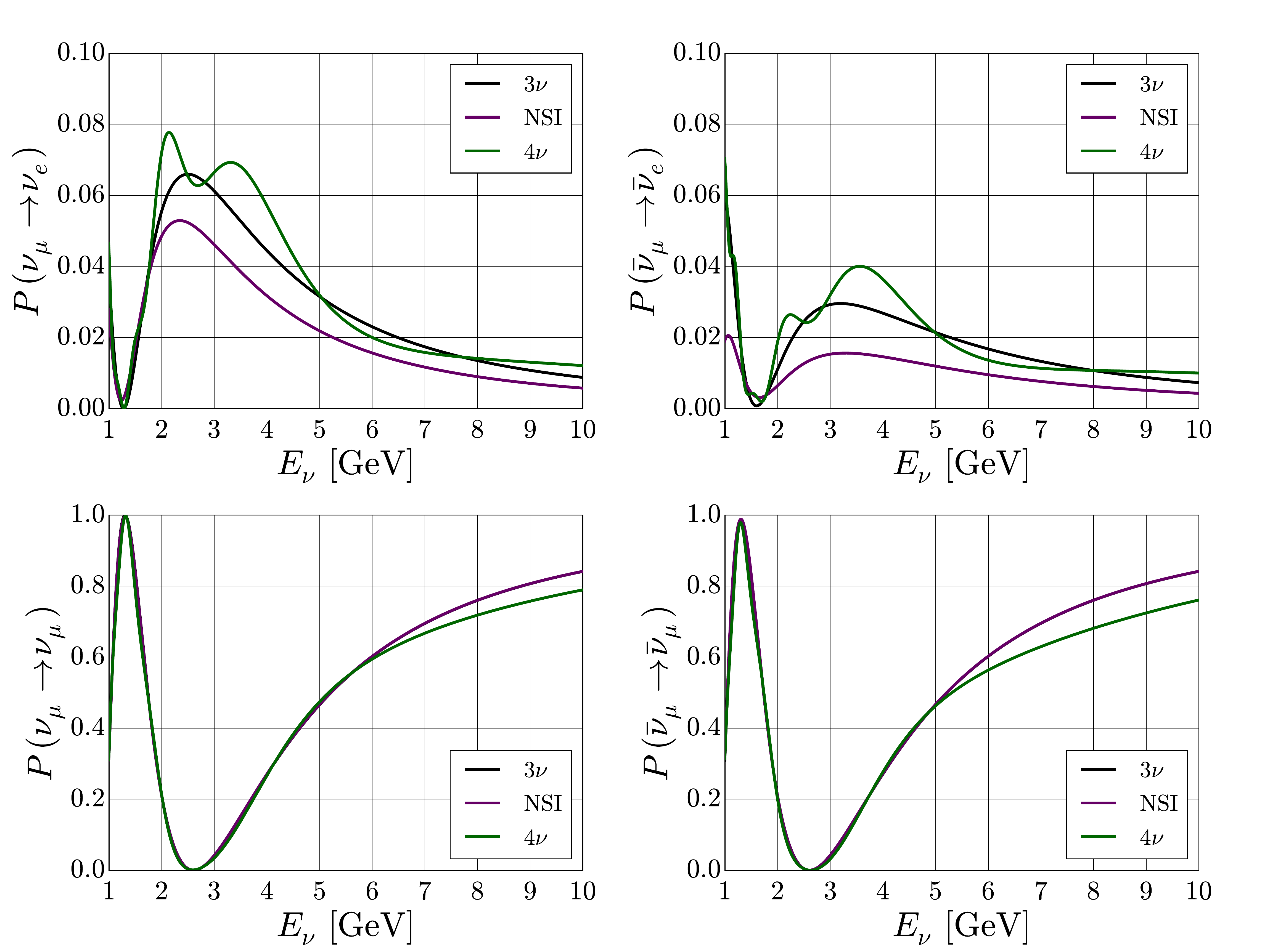}
\caption{Oscillation probabilities for a three-neutrino scenario (black, ``$3\nu$''), an NSI scenario (purple, ``NSI''), and a four-neutrino scenario (green, ``$4\nu$'') for a baseline of $1300$ km. Parameters used are $\sin^2\theta_{12} = 0.304$, $\sin^2\theta_{13} = 0.0219$, $\sin^2\theta_{23} = 0.514$, $\Delta m_{12}^2 = 7.53\times 10^{-5}$ eV$^2$, $\Delta m_{13}^2 = 2.53\times 10^{-3}$ eV$^2$, and $\delta = 0$. The NSI scenario additionally includes $\epsilon_{ee} = 2$ and $\epsilon_{e\tau} = 0.4e^{i\pi}$, while in the four-neutrino scenario we set $\phi_{ij}=\theta_{ij}$ above, for $ij= 12$, $13$, $23$, in addition to $\sin^2\phi_{14} = \sin^2\phi_{24} = 0.02$ and $\Delta m_{14}^2 = 10^{-2}$ eV$^2$.}
\label{fig:OscProbs}
\end{figure}

\begin{figure}[ht]
\centering
\includegraphics[width=0.8\linewidth]{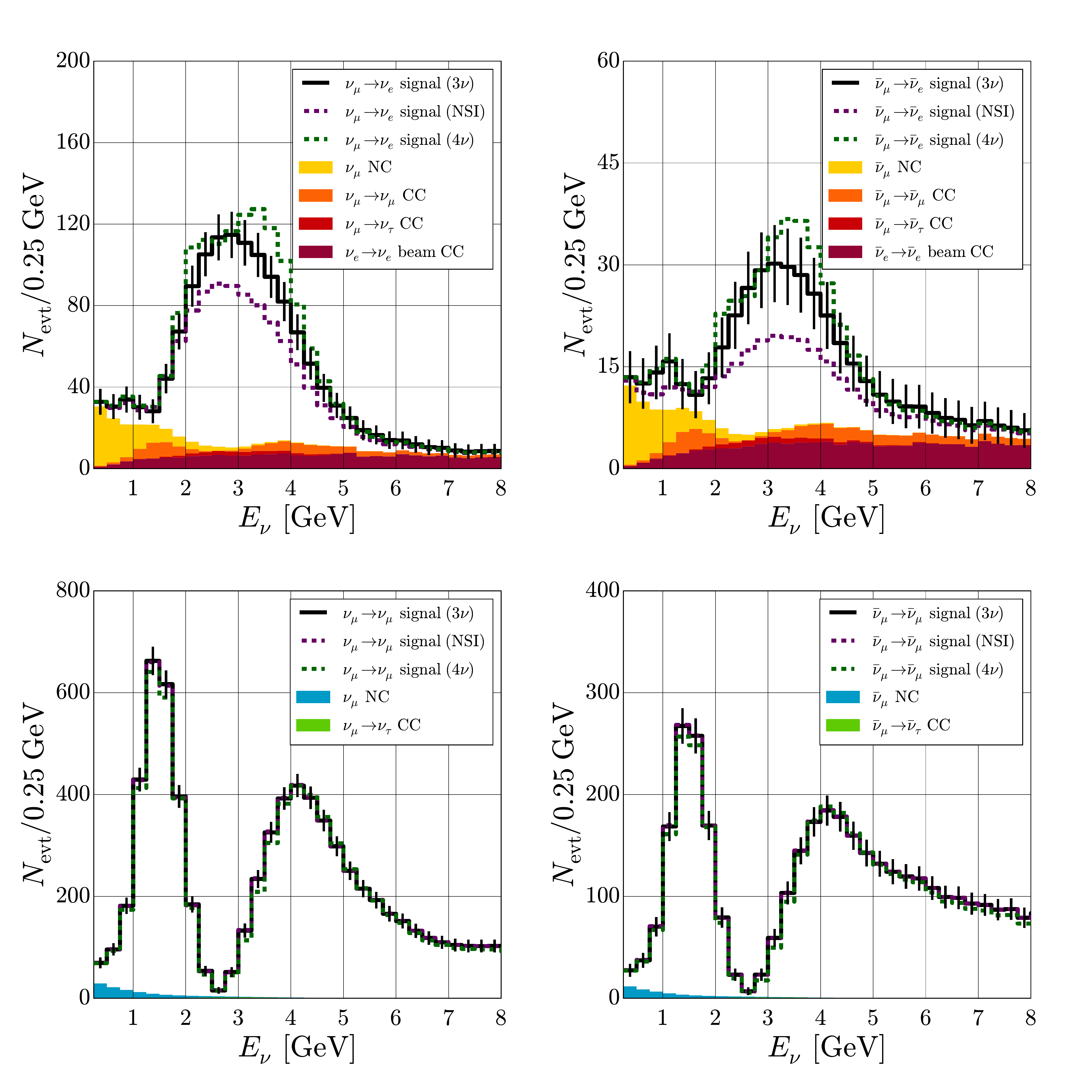}
\caption{Expected event yields at DUNE assuming a 34 kiloton detector, 1.2 MW proton beam, and three years of data collection. The top row displays yields for appearance channels $P_{\mu e}$ (left) and $P_{\bar{\mu} \bar{e}}$ (right), and the bottom row displays yields for disappearance channels $P_{\mu\mu}$ (left) and $P_{\bar{\mu}\bar{\mu}}$ (right). Three-neutrino yields are shown in black with error bars denoting statistical uncertainties, with parameters $\sin^2\theta_{12} = 0.304$, $\sin^2\theta_{13} = 0.0219$, $\sin^2\theta_{23} = 0.514$, $\Delta m_{12}^2 = 7.53\times 10^{-5}$ eV$^2$, $\Delta m_{13}^2 = 2.53\times 10^{-3}$ eV$^2$, and $\delta = 0$. Yields assuming NSI exist are shown in the dashed purple lines with the same parameters as above, along with $\epsilon_{ee} = 2$ and $\epsilon_{e\tau} = 0.4e^{i\pi}$. Yields assuming a four-neutrino scenario are shown in dashed green lines with $\sin^2\phi_{ij} = \sin^2\theta_{ij}$ from the three-neutrino oscillation parameters for $ij = 12,~13,~23$, the same values of $\Delta m_{12}^2$ and $\Delta m_{13}^2$ as above, $\eta_1 = \eta_2 = \eta_3 = 0$, $\sin^2\phi_{14} = \sin^2\phi_{24} = 0.02$, and $\Delta m_{14}^2 = 10^{-2}$ eV$^2$. The parameters used here are identical to those used in Fig.~\ref{fig:OscProbs}. Backgrounds are shown under each curve and are explained in the text.}
\label{fig:EvtYields}
\end{figure}

With the expected fluxes reported in Ref.~\cite{Adams:2013qkq} and neutrino-nucleon scattering cross-sections from Ref.~\cite{Formaggio:2013kya}, we calculate expected event yields for (anti)neutrino appearance -- using ($P_{\bar{\mu}\bar{e}}$)$P_{\mu e}$ -- and (anti)neutrino disappearance -- using ($P_{\bar{\mu}\bar{\mu}}$)$P_{\mu\mu}$ --  assuming a set of input parameters. Oscillation probabilities are computed numerically using the Hamiltonians described in Section~\ref{sec:NSIand3P1}. Fig.~\ref{fig:OscProbs} depicts the oscillation probabilities for a three-neutrino scenario, an NSI scenario, and a four-neutrino scenario. The oscillation parameters are listed in the figure caption. 
In addition to expected signal yields for these four channels, we calculate background contributions, considering four possibilities: neutral-current interactions with muon neutrinos ($\nu_\mu$ NC), charged-current interactions from unoscillated electron-type neutrinos ($\nu_e\to\nu_e$ beam CC), and charged-current interactions from muon-type ($\nu_\mu \to \nu_\mu$ CC) or tau-type ($\nu_\mu \to \nu_\tau$ CC) neutrinos. Estimated rates for these backgrounds are taken from Ref.~\cite{Adams:2013qkq}.\footnote{Updated background efficiencies are given in Ref.~\cite{Acciarri:2015uup}, and these updated efficiencies project lower backgrounds at DUNE, so the results we present should be viewed as conservative.} Fig.~\ref{fig:EvtYields} depicts the expected yields for all four channels under the same three hypotheses as those discussed in Fig.~\ref{fig:OscProbs}, including backgrounds.

In agreement with Ref.~\cite{Adams:2013qkq}, we consider energy reconstruction efficiencies of $\sigma$ [GeV] = $15\%/\sqrt{E_\nu\mathrm{\ [GeV]}}$ for electrons and taus and $\sigma$ [GeV] = $20\%/\sqrt{E_\nu \mathrm{\ [GeV]}}$ for muons. For both appearance and disappearance channels, we divide up the event yields into bins of $0.25$ GeV width between $0.5$ and $8.0$ GeV, leading to $30$ independent bins. We also include signal and background normalization uncertainties of $1\%$ and $5\%$, respectively.

By generating yields for the four channels assuming a particular hypothesis $\vec{\vartheta}_0$, we can calculate a chi-squared function\footnote{Here we assume Gaussian statistics in each bin. We find this to be a good approximation since the minimum count we expect in a bin is roughly $10$ (see Fig.~\ref{fig:EvtYields}).} analyzing the simulated data assuming any hypothesis $\vec{\vartheta}$: 
\begin{equation}
\chi^2 = \sum_{\text{channels}} \sum_{i\in\text{bins}} \frac{\left( (N_i^{(s)}(\vec{\vartheta_0}) + N_i^{(b)}(\vec{\vartheta_0})) - (\alpha \mu_i^{(s)}(\vec{\vartheta}) + \beta \mu_i^{(b)}(\vec{\vartheta})) \right)^2}{2\left(\mu_i^{(s)}(\vec{\vartheta}) + \mu_i^{(b)}(\vec{\vartheta})\right)} + \frac{\left(\alpha-1\right)^2}{2\sigma_\alpha^2} + \frac{\left(\beta-1\right)^2}{2\sigma_\beta^2},
\end{equation}
where $N_i^{(s)}(\vec{\vartheta_0})$ and $N_i^{(b)}(\vec{\vartheta_0})$ are the expected signal and background event yields in bin $i$, respectively, assuming a physical hypothesis $\vec{\vartheta_0}$, $\mu_i^{(s)}(\vec{\vartheta})$ and $\mu_i^{(b)}(\vec{\vartheta})$ are the expected signal and background event yields assuming test hypothesis $\vec{\vartheta}$, and $\alpha$ and $\beta$ are signal and background nuisance parameters with uncertainties $\sigma_\alpha$ and $\sigma_\beta$, respectively. Additionally, we include Gaussian priors on the values of $|U_{e2}|^2$ and $\Delta m_{12}^2$, which are best constrained by solar neutrino experiments and KamLAND. For the Gaussian widths, we use the values listed in the Particle Data Book~\cite{Agashe:2014kda}. We use the Markov Chain Monte Carlo program {\sc emcee} to calculate posterior likelihood distributions for the parameters of the fit hypothesis $\vec{\vartheta}$~\cite{ForemanMackey:2012ig}. See, for example,  Refs.~\citep{Berryman:2015nua,deGouvea:2015ndi,Berryman:2016szd} for more detail on this type of analysis utilizing {\sc emcee}.

\section{Results}
\label{sec:Results}

We are interested in exploring whether there are CP-conserving new physics scenarios that, if interpreted in terms of the three-neutrino-paradigm, would lead one to conclude, erroneously, that CP-invariance is violated in the lepton sector. Concretely, we simulate data assuming a particular new physics hypothesis -- NSI in Section~\ref{subsec:NSIResults} and a four-neutrino scenario in Section~\ref{subsec:SterileResults} -- and analyze them assuming the three-neutrino paradigm. We find that, for certain choices of new physics parameters, one obtains a very good fit to the wrong hypothesis while the measured best-fit value of the CP-odd parameter is consistent with maximal (i.e.~$\delta = \pm \pi/2$). Furthermore, the extracted values of  all of the other three-neutrino parameters are consistent with their current measurements~\cite{Agashe:2014kda} indicating that more information will be required in order to differentiate some forms of CP-conserving new physics from a CP-invariance violating version of the three-neutrino paradigm.

\subsection{CP-Conserving NSI}
\label{subsec:NSIResults}

\begin{table}[ht]
\begin{center}
\begin{tabular}{|c||c|c|c|c|c|c|c|c|c|c|c|c|}\hline
Parameter & $\sin^2\theta_{12}$ & $\sin^2\theta_{13}$ & $\sin^2\theta_{23}$ & $\Delta m_{12}^2$ [eV$^2$] & $\Delta m_{13}^2$ [eV$^2$] & $\delta$ & $\epsilon_{ee}$ & $\epsilon_{e\mu}$ & $\epsilon_{e\tau}$ & $\epsilon_{\mu\mu}$ & $\epsilon_{\mu\tau}$ & $\epsilon_{\tau\tau}$ \\ \hline
Value & $0.304$ & $2.19\times 10^{-2}$ & $0.514$ & $7.53\times 10^{-5}$ & $2.50\times 10^{-3}$ & $0$ & $0.730$ & $0$ & $0$ & $0$ & $0$ & $0$\\ \hline
\end{tabular}
\end{center}
\caption{Input parameters for the NSI model discussed in detail in Section~\ref{subsec:NSIResults}.}
\label{table:NSIParams}
\end{table}

\begin{figure}[ht]
\begin{center}
\includegraphics[width=0.35\linewidth]{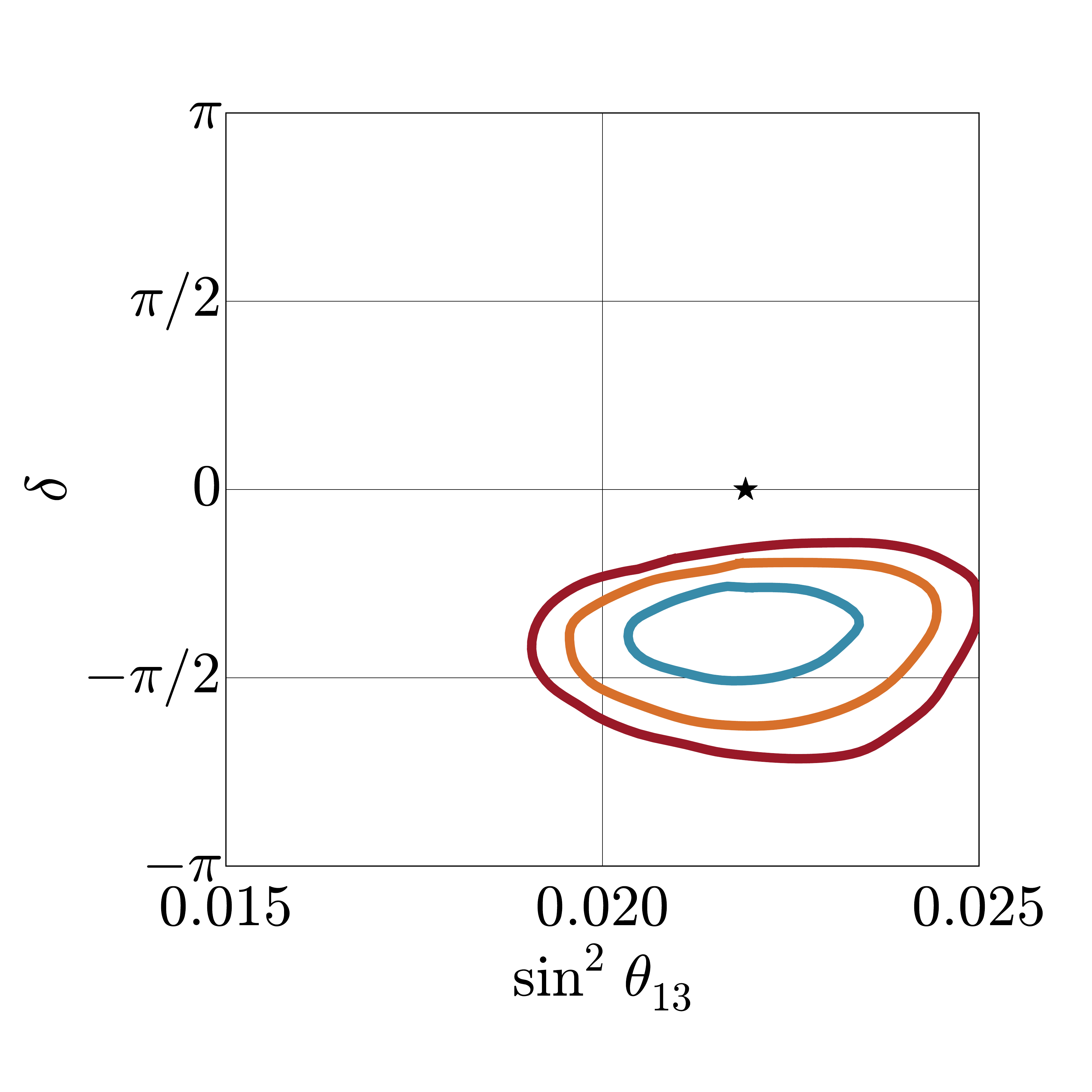}
\caption{Results of a fit assuming the three-neutrino paradigm to data simulated assuming CP-conserving NSI, with model parameters listed in Table~\ref{table:NSIParams}. All unseen parameters are marginalized. Contours shown correspond to $68.3\%$ CL (blue), $95\%$ CL (orange), and $99\%$ CL (red). Simulations utilize {\sc emcee}. The star denotes the imput values of these parameters.}
\label{fig:NSIFit}
\end{center}
\end{figure}

Table~\ref{table:NSIParams} defines a CP-conserving NSI model (note $\delta =0$), where only $\epsilon_{ee}$ is nonzero. We simulate data, as discussed in Section~\ref{sec:DUNE}, consistent with this model, and perform a fit to these simulated data assuming the three-neutrino paradigm. We obtain a very good fit; $\chi^2_\text{min}/$ dof $\simeq 120/114$.\footnote{The model contains six free parameters: $\theta_{12}$, $\theta_{13}$, $\theta_{23}$, $\Delta m_{12}^2$, $\Delta m_{13}^2$, and $\delta$, so the number of degrees of freedom in the fit (dof) is $120 - 6 = 114$.} Fig.~\ref{fig:NSIFit} depicts the measured values of $\sin^2\theta_{13}$ and $\delta$ in the  $\sin^2\theta_{13}\times \delta$ plane, at different confidence levels. Not only is a nonzero best-fit value for $\delta$ (close to $-\pi/2$) obtained, but the CP-conserving scenarios $\delta = 0$ and $\delta = \pi$ are excluded at over 99\% CL. The extracted values of the other oscillation parameters, with one-sigma error estimates, are listed in Table~\ref{table:NSIFitResults}. They are consistent with the values in Table~\ref{table:NSIParams} and consistent with the current values extracted from existing data, obtained assuming the three-neutrino paradigm is correct. If nature is consistent with Table~\ref{table:NSIParams}, DUNE data will very likely be interpreted as evidence that the three-neutrino paradigm is correct, and that CP-invariance is strongly violated in the lepton sector. 

\begin{table}[ht]
\begin{center}
\begin{tabular}{|c|c|c|c|c|c|c|}
\hline
Parameter & $\sin^2\theta_{12}$ & $\sin^2\theta_{13}$ & $\sin^2{\theta_{23}}$ & $\Delta m_{12}^2$ [eV$^2$] & $\Delta m_{13}^2$ [eV$^2$] & $\delta$ \\ \hline
Measurement & $0.310 \pm 0.016$ & $(2.20^{+0.12}_{-0.09})\times 10^{-2}$ & $0.514\pm 0.016$ & $(7.52 \pm 0.19)\times 10^{-5}$ & $(2.50 \pm 0.01)\times 10^{-3}$ & $-1.21^{+0.24}_{-0.29}$ \\ \hline
\end{tabular}
\end{center}
\caption{Measurement of three-neutrino parameters assuming data consistent with nonzero NSI from Table~\ref{table:NSIParams}, analyzed assuming the three-neutrino paradigm. The ranges displayed correspond to 68.3\% CL. Results are consistent with expectations at DUNE and current results in Ref.~\cite{Agashe:2014kda}.}
\label{table:NSIFitResults}
\end{table}

The scenario in Table~\ref{table:NSIParams} is a CP-conserving version of the three-neutrino paradigm, plus a large $\epsilon_{ee}$. A nonzero $\epsilon_{ee}$ is degenerate with the value of the Earth's density, so a large $\epsilon_{ee}$ can be mimicked by judiciously modifying the matter density traversed by the neutrinos. The result above can be reinterpreted in the following way: a large -- almost a factor of two -- underestimation of the Earth's density along the path of the neutrinos will lead one to incorporate the wrong matter effects in the data analysis and incorrectly conclude that CP-invariance is violated. Related results can be found in the literature (e.g. Ref.~\cite{Ohlsson:2003ip}) in studies of the impact of uncertainties in the matter density when measuring $\delta$.

\begin{figure}[ht]
\centering
\includegraphics[width=0.6\linewidth]{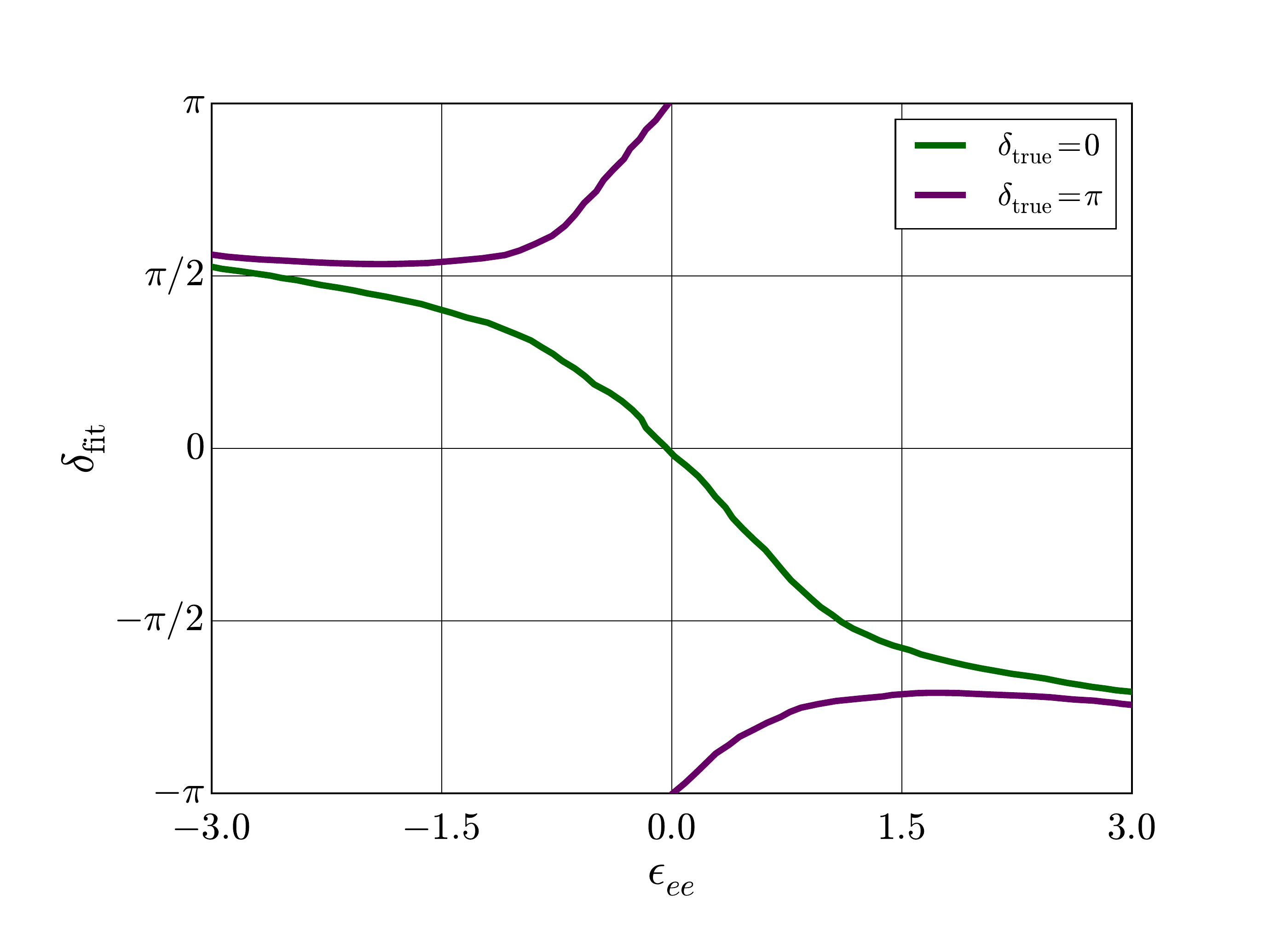}
\caption{$\delta_{\rm fit}$, the extracted value of $\delta$ assuming a three-neutrino hypothesis, as a function of the input value of $\epsilon_{ee}$, for input values of $\delta_{\text{true}} = 0$ (green) or $\pi$ (purple).}
\label{fig:DeltaFitted}
\end{figure}

Fig.~\ref{fig:DeltaFitted} illustrates this phenomenon more quantitatively. We repeat the exercise above, for physical values of $\delta_\text{true} = 0$ (green) and $\pi$ (purple) -- the two CP-conserving cases -- and different values of $\epsilon_{ee}$.  Fig.~\ref{fig:DeltaFitted} depicts the extracted value of $\delta$, $\delta_{\rm fit}$, as a function of $\epsilon_{ee}$, obtained under the hypothesis that the three-flavor paradigm is correct (i.e., fixing $\epsilon_{ee}=0$). For CP-conserving $\epsilon_{ee}$ NSI, the extracted incorrect value of $\delta$ can be anything.

\begin{figure}[!hpt]
\centering
\includegraphics[width=0.8\linewidth]{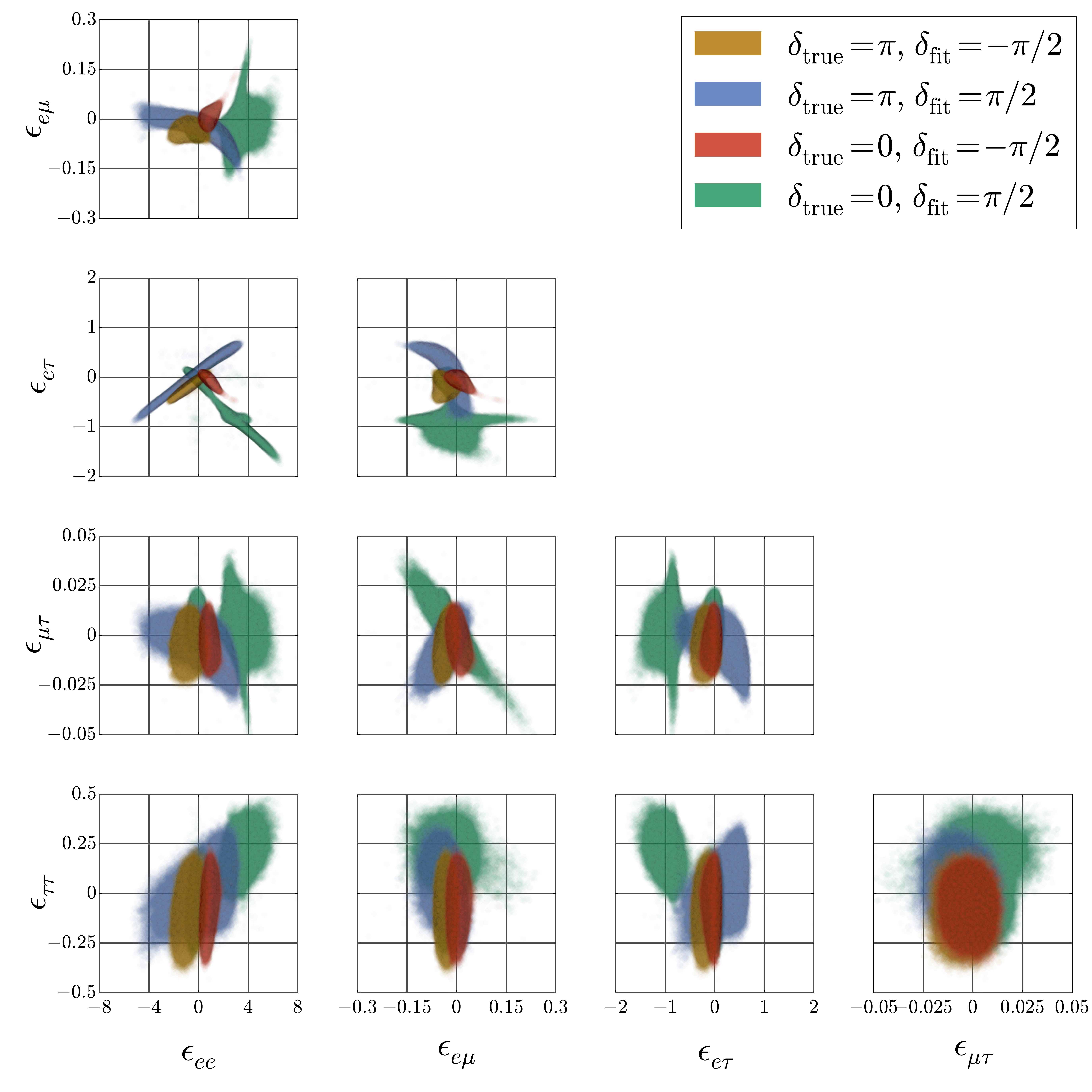}
\caption{Regions of CP-conserving NSI parameter space with a physical value of $\delta_{\text{true}}$ that mimics a three-neutrino scenario with $\delta = \delta_{\text{fit}}$. Three-neutrino parameters, not shown, are consistent with values from Ref.~\cite{Agashe:2014kda}. Since the NSI parameters are CP-conserving, and therefore real, we allow all $\epsilon_{\alpha\beta}$ to be negative, hence including both $\phi_{\alpha\beta} = 0,\ \pi$.}
\label{fig:NSISpace}
\end{figure}

The ability to fake the three-neutrino paradigm with CP-invariance violation extends beyond NSI scenarios with a nonzero $\epsilon_{ee}$. We find that the capability of CP-conserving NSI scenarios to mimic CP-invariance violation with no non-standard interactions is generic. It can occur for either the normal ($\Delta m_{13}^2 > 0$) or inverted ($\Delta m_{13}^2 < 0$) mass hierarchy, when the true value of $\delta$ is $0$ or $\pi$ (i.e.\ $U_{e3} = s_{13} e^{i\delta} < 0$). We also find that the three-neutrino hypothesis may prefer fake $\delta$ values large and positive (close to $\pi/2$) or large and negative (close to $-\pi/2$). Fig.~\ref{fig:NSISpace} depicts, for different scenarios, regions of CP-conserving NSI parameter space consistent with a maximal CP-invariance violating three-neutrino hypothesis, assuming the normal mass hierarchy. Here, we define two hypotheses as consistent when $\chi^2 /$ dof $\lesssim 134/114$ for the wrong hypothesis, or roughly, when the wrong hypothesis is ruled out at less than $1.7\sigma$. 

\subsection{CP-Conserving Four-Neutrino Model}
\label{subsec:SterileResults}

\begin{table}[ht]
\begin{center}
\begin{tabular}{|c||c|c|c|c|c|c|c|c|c|c|c|c|}
\hline
Parameter & $\sin^2\phi_{12}$ & $\sin^2\phi_{13}$ & $\sin^2\phi_{23}$ & $\sin^2\phi_{14}$ & $\sin^2\phi_{24}$ & $\sin^2\phi_{34}$ & $\Delta m_{12}^2$ [eV$^2$] & $\Delta m_{13}^2$ [eV$^2$] & $\Delta m_{14}^2$ [eV$^2$] & $\eta_1$ & $\eta_2$ & $\eta_3$ \\ \hline
Value & $0.540$ & $3.84\times 10^{-2}$ & $0.531$ & $0.404$ & $4.2\times 10^{-4}$ & $0$ & $7.50\times 10^{-5}$ & $2.48\times 10^{-3}$ & $3.87\times 10^{-3}$& $0$ & $\pi$ & $0$ \\ \hline
\end{tabular}
\end{center}
\caption{Input parameters for the four-neutrino scenario discussed in detail in Section~\ref{subsec:SterileResults}. While $\sin^2\phi_{12}$ is much larger than the accepted value for $\sin^2\theta_{12} \simeq 0.3$, the matrix element $U_{e2}$ is roughly the same in both scenarios.}
\label{table:SterileParams}
\end{table}

\begin{figure}[ht]
\begin{center}
\includegraphics[width=0.35\linewidth]{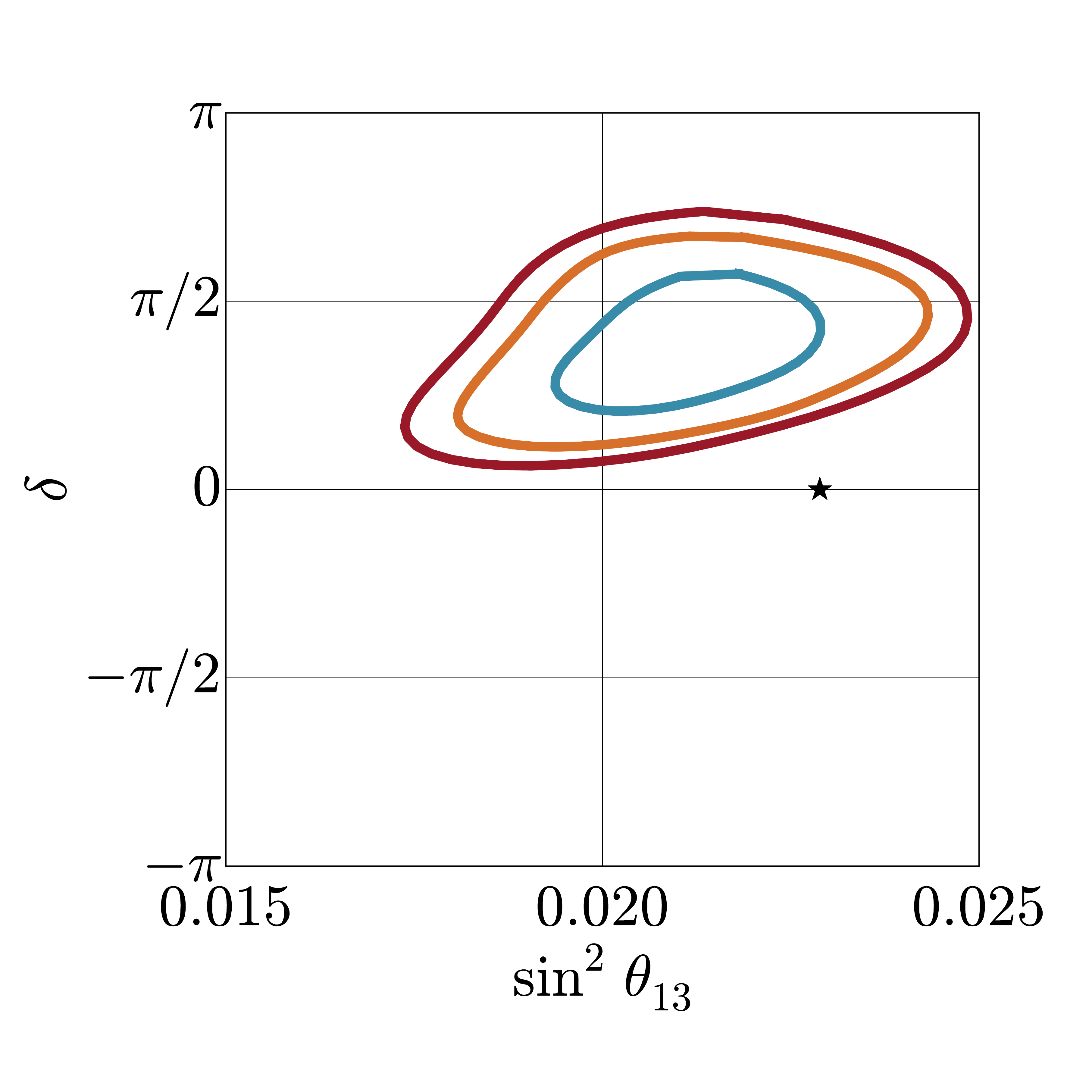}
\caption{Results of a three-neutrino-paradigm fit to data simulated assuming a four-neutrino scenario, with parameters listed in Table~\ref{table:SterileParams}. All unseen parameters are marginalized. Contours shown correspond to $68.3\%$ CL (blue), $95\%$ CL (orange), and $99\%$ CL (red). Simulations utilize {\sc emcee}. The star denotes the input values of these parameters, where we identify $\sin^2\theta_{13} = |U_{e3}|^2 =  \sin^2\phi_{13} \cos^2\phi_{14}$, cf Eq.~(\ref{eq:Ue3}).}
\label{fig:SterileFit}
\end{center}
\end{figure}

Table~\ref{table:SterileParams} defines a CP-conserving four-neutrino model (note all $\eta_{i}$, $i=1,$ $2,$ $3$, are zero or $\pi$).\footnote{Strictly speaking, we define as CP-conserving those scenarios where there is no fundamental CP-invariance violation as far as $\nu_{\mu}\to\nu_e$ oscillations are concerned. If there is large CP-invariance violation in the tau-sector, our analyses would be insensitive to it.} As in Section~\ref{subsec:NSIResults}, we simulate data consistent with this four-neutrino hypothesis and analyze them assuming the three-neutrino paradigm -- again, a fit with $114$ degrees of freedom. We obtain a very good fit; $\chi^2_\text{min}/$ dof $\simeq 120/114$. Fig.~\ref{fig:SterileFit} depicts the measured values of $\sin^2\theta_{13}$ and $\delta$ in the  $\sin^2\theta_{13}\times \delta$ plane, at different confidence levels. The best-fit  value of $\delta$ is large (close to $\pi/2$), and $\delta = 0$ and $\delta = \pi$ are excluded at over 99\% CL. While the values $\sin^2\phi_{12,13,23}$ are not equal to the current best-fit values of $\sin^2\theta_{12,13,23}$, the measured ranges of the latter, listed in Table~\ref{table:SterileFitResults}, are consistent with current oscillation data. If nature is consistent with Table~\ref{table:SterileParams}, DUNE data will very likely be interpreted as evidence that the three-neutrino paradigm is correct, and that CP-invariance is strongly violated in the lepton sector.     

\begin{table}[ht]
\begin{center}
\begin{tabular}{|c|c|c|c|c|c|c|}
\hline
Parameter & $\sin^2\theta_{12}$ & $\sin^2\theta_{13}$ & $\sin^2{\theta_{23}}$ & $\Delta m_{12}^2$ [eV$^2$] & $\Delta m_{13}^2$ [eV$^2$] & $\delta$ \\ \hline
Measurement & $0.310 \pm 0.016$ & $(2.12^{+0.12}_{-0.11})\times 10^{-2}$ & $0.517^{+0.016}_{-0.019}$ & $(7.52^{+0.19}_{-0.20})\times 10^{-5}$ & $(2.51 \pm 0.01)\times 10^{-3}$ & $1.169^{+0.41}_{-0.39}$ \\ \hline
\end{tabular}
\end{center}
\caption{Measurement of three-neutrino parameters assuming data consistent with a fourth neutrino and parameters from Table~\ref{table:SterileParams}, analyzed assuming the three-neutrino paradigm. The ranges displayed correspond to 68.3\% CL. Results are consistent with expectations at DUNE and current results in Ref.~\cite{Agashe:2014kda}.}
\label{table:SterileFitResults}
\end{table}

\begin{figure}[ht]
\centering
\includegraphics[width=0.5\linewidth]{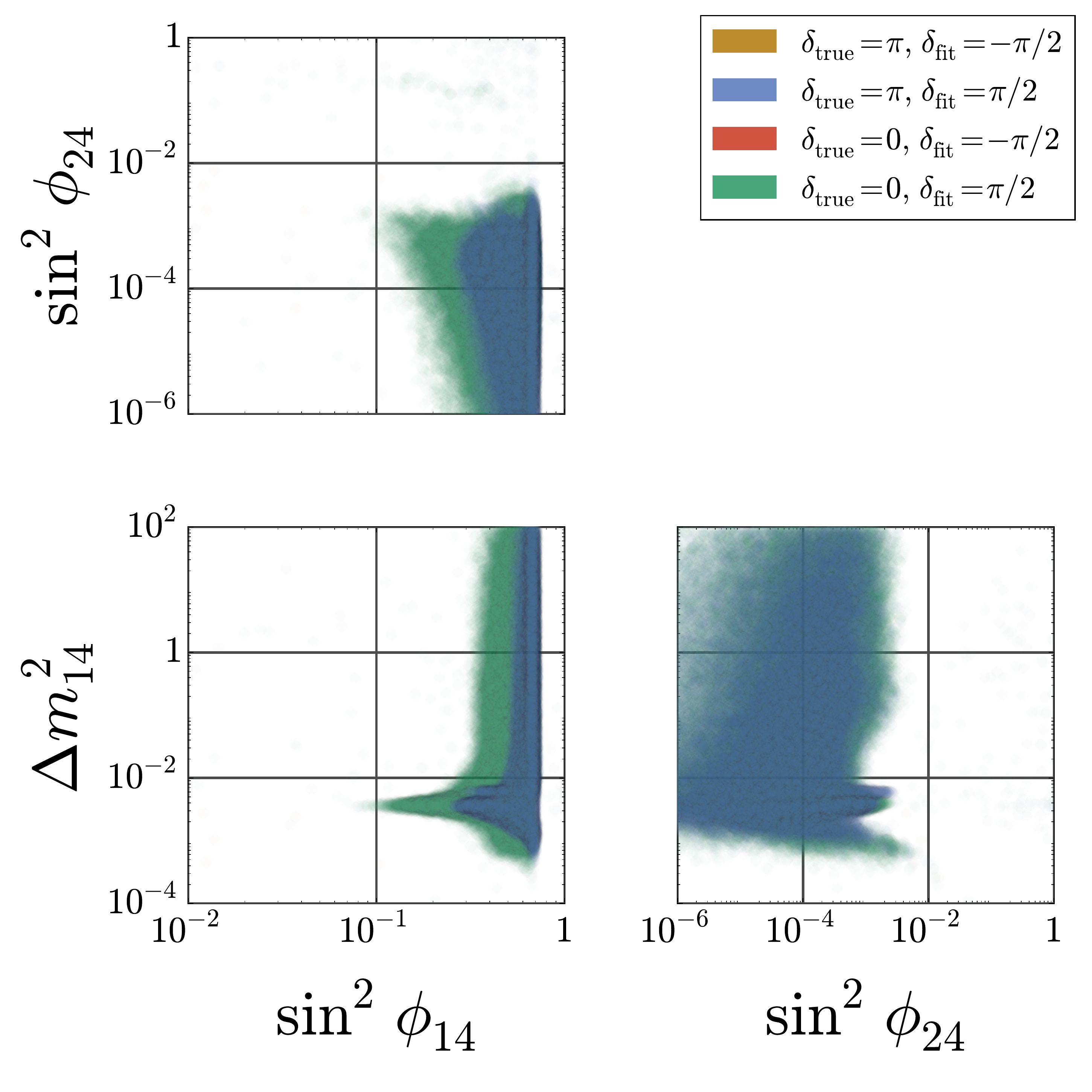}
\caption{Regions of CP-conserving four-neutrino parameter space with a physical value of $\eta_1 = \delta_{\text{true}}$ that mimics a three-neutrino scenario with $\delta = \delta_{\text{fit}}$. Three-neutrino parameters, not shown, are consistent with values from Ref.~\cite{Agashe:2014kda}. The new matrix elements $U_{e4}$ and $U_{\mu 4}$ are allowed to be positive or negative, i.e. $\eta_2 - \eta_3 = 0$ or $\pi$.}
\label{fig:SterileSpace}
\end{figure}

As with the NSI case in Sec.~\ref{subsec:NSIResults}, we find that there are several CP-conserving four-neutrino scenarios that mimic the three-neutrino paradigm with large CP-invariance violation at DUNE. Fig.~\ref{fig:SterileSpace} depicts, for different scenarios, regions of CP-conserving four-neutrino parameter space consistent with a maximal CP-invariance violating three-neutrino hypothesis, assuming the normal mass hierarchy. Again, we define two hypothesis as consistent when $\chi^2 /$ dof $\lesssim 134/114$ for the wrong hypothesis, or roughly, when the wrong hypothesis is ruled out at less than $1.7\sigma$. When it comes to inducing fake CP-violation, four-neutrino scenarios are not as efficient as NSI scenarios. We do not find, for example, CP-conserving points that are fit incorrectly by a three-neutrino hypothesis and $\delta = -\pi/2$ -- for the normal hierarchy, a four-neutrino scenario mimics $\delta \sim \pi/2$ more consistently than $\delta \sim -\pi/2$. The lack of red and gold points in Fig.~\ref{fig:SterileSpace} is a manifestation of this. While there is a large region of parameter space consistent with $\delta_{\text{fit}}=\pi/2$, we note that a large portion of it is excluded by past and present searches for a fourth neutrino, e.g. Daya Bay~\cite{An:2014bik}, Bugey~\cite{Declais:1994su}, MINOS~\cite{Timmons:2015lga}, and IceCube~\cite{TheIceCube:2016oqi}. Next-generation short-baseline searches for sterile neutrinos are also likely to make a discovery or significantly constrain parts of the parameter space for large values of $\Delta m^2_{14}$ before DUNE is scheduled to take data, see for example, Refs.~\cite{Abazajian:2012ys,Antonello:2015lea,Ashenfelter:2015uxt}. 

\section{Discussion and Conclusions}
\label{sec:Conclusions}

We have shown there is the possibility that (a) the leptonic sector is CP-conserving and (b) there are new, CP-conserving neutrino--matter interactions or new light neutrino states, but,  nonetheless, all current neutrino-oscillation data and future DUNE data will be consistent with the standard three-massive-neutrinos paradigm and large CP-invariance violation in the lepton sector. Figures \ref{fig:NSISpace} and \ref{fig:SterileSpace} illustrate that this does not occur only for special choices of the new-physics parameters: there are plenty of CP-conserving, new-physics scenarios that mimic the three-neutrino paradigm with large CP-invariance violation. 

Matter effects are mostly responsible for the confusion. The matter background through which neutrinos propagate is CPT-violating and renders, even in the absence of CP-invariance violating fundamental physics effects, neutrino oscillations different from antineutrino oscillations. This implies that one needs to model the effects of matter in order to extract out the fundamental CP-violating effects. Hence, it is not too surprising that if one uses the wrong model to handle matter effects, spurious CP-invariance violating effects might arise. In summary, while matter effects are invaluable for neutrino oscillation research -- they allow powerful sensitivity to the mass-hierarchy and provide unique sensitivity to new phenomena -- they are a nuisance when it comes to determining whether CP-invariance is violated in the lepton sector. 

It is easy to see that matters effects are to blame when it comes to the NSI examples discussed here. If one were to turn off the effects of matter, all NSI effects become unobservable (given the assumptions we make here) and, of course, there would be no fake CP-violation due to new physics. We have also investigated whether matter effects were mostly responsible for the fake CP-violation in the four-neutrino scenarios discussed here. We did this by asking whether it is possible to fake, at DUNE, large three-neutrino CP-violation with a CP-conserving, four-neutrino scenario assuming the neutrinos propagate in vacuum. The answer, it turns out, is no. 
 
This diagnosis allows one to identify how to disentangle these CP-conserving new-physics scenarios from the standard three-neutrino paradigm with large CP-violation. Since the scenarios we picked are such that the only mis-measured parameter, to leading order, is $\delta$, not much help is expected from the current data, which is mostly insensitive to CP-invariance violating effects, or from next-generation experiments sensitive to the solar parameters, like the Jiangmen Underground Neutrino Observatory (JUNO) \cite{An:2015jdp}. It turns out that, in the cases considered here, the $\nu_{\mu}$ disappearance channel does not play a significant role, so precision measurements of the $\nu_{\mu}$ disappearance at different $L$ and $E_{\nu}$ are also not expected to help resolve the degeneracy significantly, except, perhaps, for atmospheric neutrinos. 
 
Experiments aimed at observing $\nu_{\mu}\to\nu_e$ (and the CP-conjugated channel) with good enough precision and high enough statistics so $\delta$-driven CP-violating effects can be observed, on the other hand, should prove invaluable. This is especially true when these contain $L/E_{\nu}$ values similar to those available to DUNE but have access to very distinct values of $L$ and $E_{\nu}$. An obvious candidate is the Hyper-Kamiokande long-baseline experiment \cite{Kearns:2013lea,Abe:2014oxa,Abe:2015zbg}. Other possibilities include  high-statistics, high-precision measurements of the atmospheric neutrino flux \cite{Akhmedov:2012ah,Esmaili:2013fva,Choubey:2014iia,Mocioiu:2014gua,Ohlsson:2013epa}, like the Precision IceCube Next Generation Upgrade (PINGU) \cite{Aartsen:2014oha} or potential upgrades. Very recently, a study of NSI effects and high-energy, astrophysical neutrinos also became public \cite{Gonzalez-Garcia:2016gpq}.
 
\begin{figure}[ht]
\centering
\includegraphics[width=\linewidth]{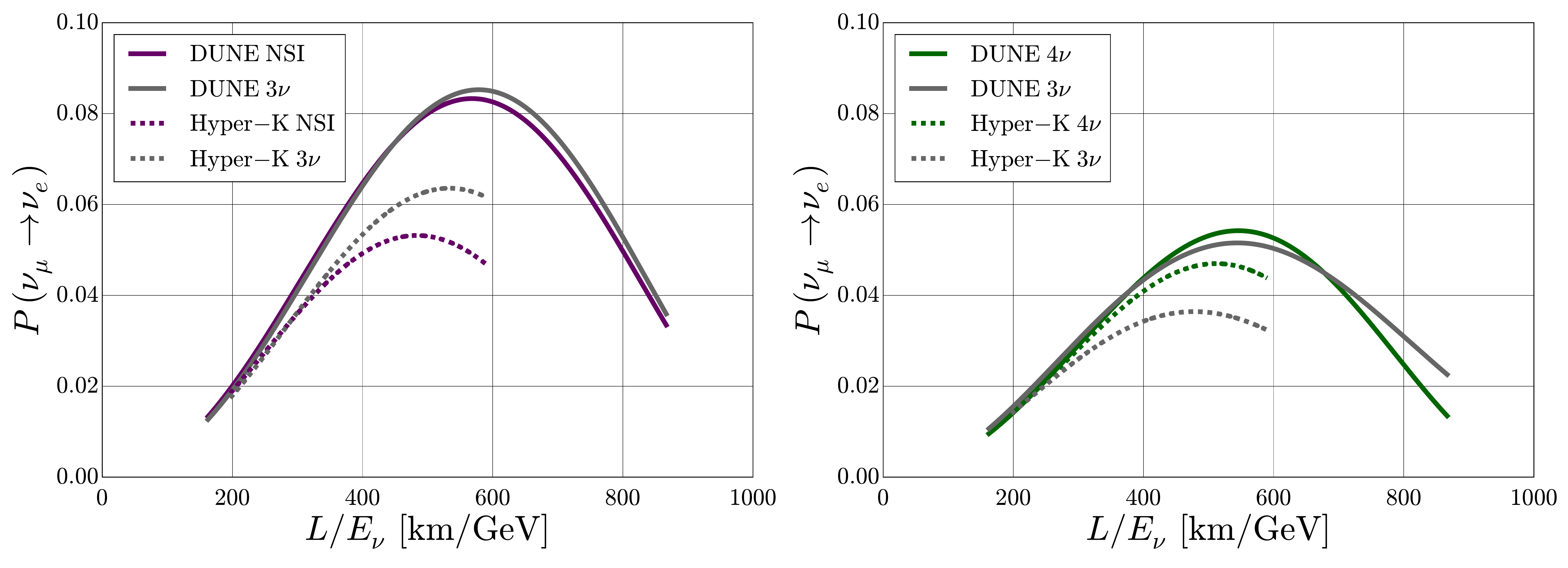}
\caption{$P_{\mu e}$ as a function of $L/E_{\nu}$ for $L=1300$~km (solid, DUNE) and $L=295$~km (dotted, Hyper-K), for (left) the NSI model in Table~\ref{table:NSIParams} and the corresponding best-fit three-neutrino model in Table~\ref{table:NSIFitResults} and (right) the four-neutrino model in Table~\ref{table:SterileParams} and the corresponding best-fit three-neutrino model in Table~\ref{table:SterileFitResults}. The upper bounds to the Hyper-K curves correspond to lowest-energy neutrinos produced in the experiment.}
\label{fig:DUNEHKComp}
\end{figure}

Figure~\ref{fig:DUNEHKComp} depicts $P_{\mu e}$ as a function of $L/E_{\nu}$ for $L=1300$ km (DUNE) and $L=295$~km (Hyper-K), for (left) the NSI model in Table~\ref{table:NSIParams} and the corresponding best-fit three-neutrino model discussed in Section~\ref{subsec:NSIResults} -- see Table~\ref{table:NSIFitResults} -- and (right) the four-neutrino model in Table~\ref{table:SterileParams} and the corresponding best-fit three-neutrino model discussed in Section~\ref{subsec:SterileResults} -- see Table~\ref{table:SterileFitResults}. For DUNE, the three-neutrino and the new-physics curves agree, of course,  very well, but the two hypotheses disagree significantly -- they differ by as much as 25\% --  for the HyperK baseline. This means that if, for example, the scenario in Table~\ref{table:NSIFitResults} is realized in nature, DUNE will point to the three-neutrino paradigm and large CP-invariance violation, while Hyper-K will point to the three-neutrino paradigm and no CP-invariance violation. The solution to this inconsistency will be the existence of new physics effects at DUNE which manifest themselved as a nonzero $\epsilon_{ee}$.

In conclusion, our understanding of neutrino properties has evolved dramatically in the last two decades. Nonetheless, while our current understanding of the leptonic sector is robust and consistent with almost all neutrino data, there is still plenty of room for new phenomena. Ambitious, next-generation long-baseline neutrino oscillation experiments are required in order to non-trivially test the three-massive-neutrinos paradigm. Such tests are not only dramatically important in their own right but are also required in order to avoid ambiguities from potentially surrounding the answers to fundamental questions, including whether there is CP-invariance violation in the lepton sector.

\begin{acknowledgments}
We are indebted to Pilar Coloma for many useful discussions and for collaboration during early stages of this work. We also thank Jeff Berryman for productive discussions. This work is supported in part by the DOE grant \#DE-FG02-91ER40684.  
\end{acknowledgments}

\bibliographystyle{apsrev-title}
\bibliography{CPBib}{}

\end{document}